\documentclass[final]{IEEEtran}

\usepackage{epsfig}
\usepackage{url}
\usepackage{amsmath}
\usepackage{multirow}
\usepackage{subfig}

\newcommand{\refFig}[1]{Figure~\ref{#1}}
\newcommand{\refTab}[1]{Table~\ref{#1}}
\newcommand{\refSec}[1]{Section~\ref{#1}}

\begin{document}

\title{BitTorrent Experiments on Testbeds: A Study of the Impact of Network Latencies}

\author{
\IEEEauthorblockN{Ashwin Rao,}
\and
\IEEEauthorblockN{Arnaud Legout, and}
\and
\IEEEauthorblockN{Walid Dabbous}\\
\IEEEauthorblockA{INRIA, France.\\
\{ashwin.rao, arnaud.legout, walid.dabbous\}@inria.fr}
}

\maketitle
\pagestyle{empty}
\thispagestyle{empty}

\begin{abstract}
In this paper, we study the impact of network latency on the
time required to download a file distributed using BitTorrent. This
study is essential to understand if testbeds can be used for
experimental evaluation of BitTorrent. We observe that
the network latency has a marginal impact on the time required to
download a file; hence, BitTorrent experiments can performed on
testbeds.
\end{abstract}

\section{Introduction}

\noindent
Testbeds such as PlanetLab and Grid5000 are widely used to study the
performance of communication protocols and networking
applications. One commonly used practice while performing experiments
on such testbeds is to run multiple instances of the application being
studied on the same machine. However, one primary shortcoming of this
approach is the absence of any network latency between the instances of
the application running on the same machine. Further, in experiments
involving more than one machine, the latency between the machines
present in the same local area network (LAN) is negligible. In this
paper we study the impact of network latency on the outcome of
experiments that are performed on testbeds to evaluate the performance
of BitTorrent.

The BitTorrent Protocol internally uses the Transmission Control
Protocol (TCP) while distributing the
content~\cite{Cohen_2008_BitTorrentProtocol}. The steady-state
throughput of TCP is function of the round-trip time
(RTT)~\cite{Mathis_1997_MBTCP}. Further, the slow start and congestion
avoidance phase of TCP introduce a \emph{ramp up period} which is
required to attain a throughput equal to the minimum of the
network throughput and the rate at which the application is sending
data. This ramp up period is a function of the RTT and the rate at
which the data is being uploaded. BitTorrent allows the users to limit
the rate at which data is uploaded; as the time duration of an upload
by a peer is in the order of seconds, we believe that the time
required to transfer pieces of a file is not affected by such
variations in the TCP throughput. Our experiments show that the RTT
(and hence the latency) between the peers in the torrent has a
marginal impact (less than $15\%$) on the time required to download a
file.

The details of the methodology and the tools used are presented
in \refSec{sec:Methodology}. We initially assume the latency between any
two peers in the torrent to be the same (\emph{homogeneous
  latency}); the impact of homogeneous latency on the
time required to download a file are presented in
\refSec{sec:Homogeneous}. The results without this assumption are
presented in \refSec{sec:Heterogeneous}, followed by the conclusions in
\refSec{sec:Conclusion}.

\section{Methodology}
\label{sec:Methodology}

In this paper we use the terminology used by the BitTorrent
community. A \emph{torrent}, also known as a BitTorrent session or a
swarm, consists of a set of peers that are interested in having a
copy of the given content. A peer in a torrent can be in two states:
the \emph{leecher} state when it is downloading the contents, and the
\emph{seed} state when it has a copy of the content being
distributed. A \emph{tracker} is a server that keeps track of the
peers present in the torrent.

\subsection{Experiment Scenarios}

We consider a torrent consisting of one tracker and a finite
number of peers; a few of these peers are seeds,
while the rest of the peers are leechers. We assume that the
peers remain in the torrent until all the leechers have finished
downloading the file.

The metric used to study the impact of the network latency between
the peers is the \emph{download completion time}, the time required to the
download the file distributed using BitTorrent. We use the following
network topologies to evaluate the impact of latency on download
completion time of a file.
\begin{enumerate}
\item \emph{Homogeneous Latency.} The latency between any two peers
  in the torrent is the same in this network topology. This topology
  provides \emph{an upper bound on the download  completion time} when
  the maximum round trip time between the peers in a torrent is
  known. Further, this setting was used to give an insight on the
  threshold of the latency between the peers beyond which the latency
  affects the download completion time.

\item \emph{Heterogeneous Latency.} The peers are grouped together to
  abstract Autonomous Systems (AS). We assume that the  latency between
  any two peers in a given AS is the same and that all ASes are fully
  meshed. Further, we assume that the inter-AS latency is greater than
  the intra-AS latency; we also assume symmetric latency in the
  upload and download links within an AS and between ASes.
\end{enumerate}

\begin{figure}
\begin{centering}
\includegraphics[width=0.8\columnwidth]{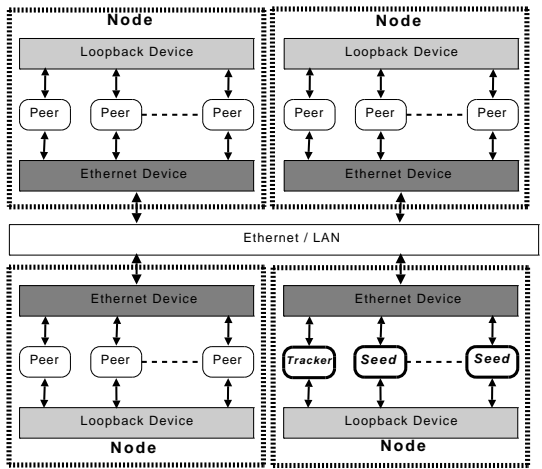}
\caption{Topology of the peers in the machines used for the
  experiment. One machine for the tracker and the initial seed, and
  three machines each with one hundred leechers.}
\label{fig:Grid5KTopology}
\end{centering}
\vspace{-0.1in}
\end{figure}

\begin{figure}
\begin{centering}
\includegraphics[width=0.8\columnwidth]{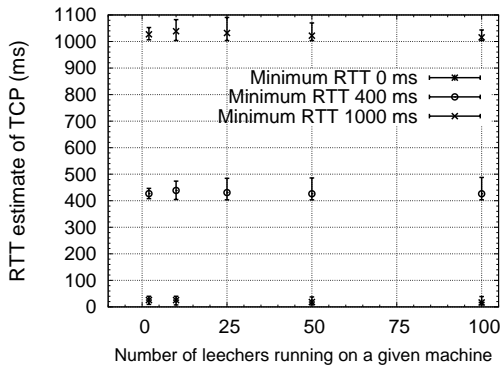}
\caption{Impact of the number of leechers running on a machine on the RTT
  estimate of TCP. Error bars indicating the $5^{th}$ and
  $95^{th}$ percentile of RTT estimated by TCP for all the peers in five
  iterations. Increasing the number of leechers running on a
  given machine has a marginal impact on the RTT estimated by TCP.}
\label{fig:ImpNumNodes}
\end{centering}
\vspace{-0.1in}
\end{figure}

\begin{figure}
\begin{centering}
\includegraphics[angle=-90,width=\columnwidth]{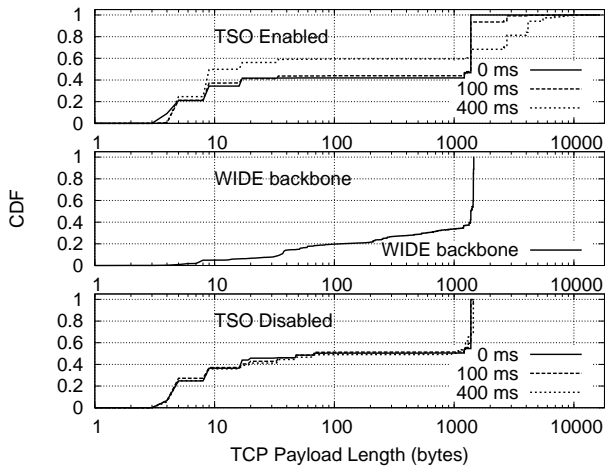}
\caption{CDF of TCP Payload Length. The maximum payload length with
  TSO enabled is much greater than that observed in the WIDE
  Backbone. Disabling TSO ensures that the maximum payload length is
  similar to that observed in the WIDE backbone.}
\label{fig:ImpTSO}
\end{centering}
\end{figure}

All the experiments were performed in a private torrent consisting of
one tracker, one initial seed (henceforth called as the seed), and 300
leechers; these experiments were carried
out on the Grid'5000 experimental testbed~\cite{grid5000}. A 50 MB
file was distributed in this torrent where the upload rates of the
leechers and the seed was varied from 10~kB/s to 100~kB/s. As
shown in~\refFig{fig:Grid5KTopology}, four machines with Linux as their
operating system were used to run the instances of the peers in the
torrent; one machine was used for the tracker and the seed, and each
of the other three machines ran 100 instances of the leechers. A pair
of peers in the torrent, including the tracker, the seed, and the
leechers, communicate with each other using either the loopback
interface or the ethernet interface. The latency between the peers was
varied from 0~ms to 500~ms using the Netem
module~\cite{Hemminger_2005_NETEM}. This latency represents the one
way delay observed by a packet, hence \emph{the round-trip time between any
two peers is at least twice the latency mentioned}. Homogeneous latency was
achieved by adding the same latency on the loopback and the ethernet
interface. Similarly, heterogeneous latency was achieved by adding a
latency on the ethernet interface of a given machine that is greater
than the latency added on the loopback interface of the same machine.

The following torrent configurations were used to vary the upload rate of
the peers:
\begin{enumerate}
\item \emph{Seed and Leechers Slow.} In this setting, the upload rate
  of the peers was limited to 10~kB/s and 20~kB/s.
\item \emph{Seed and Leechers Fast.} In this setting, the upload rate
  of the peers was limited to 50~kB/s and 100~kB/s.
\item \emph{Seed Fast and Leechers Slow.} In this setting, the upload rate
  of the initial seed was limited to 50~kB/s while the upload rate of the
  leechers was limited to 20~kB/s.
\end{enumerate}

\subsection{Testbed Configuration}

The Netem module buffers the TCP frames which are in flight for a time
period equal to the latency being emulated. A buffer size of
100000 frames was used in each machine to support up to 1000 frames of each
peer to be in flight. To ensure that the machines are
capable of running 100 peers uploading at 100~kB/s without affecting
the added latency, we varied the number of leechers running on a
machine from 4 to 100. The \verb,TCP_INFO, option for the
\verb,getsockopt, method of the socket library was used to sample the
RTT estimated by TCP each time a \verb,send, system call was issued on
a  socket. \refFig{fig:ImpNumNodes} shows the
average RTT estimate with the error bars representing the 95th and 5th
percentiles of all the peers in five iterations. We observe that the
number of leechers running a given machine has a marginal impact on
the average RTT estimated by TCP when each of the leechers has
its maximum upload rate limited to 100~kB/s.

The Maximum Transmission Unit (MTU) for the loopback interface is
typically greater than the MTU for other network interfaces such as
Wi-Fi and ethernet. To avoid the impact of large frames being
exchanged between the peers we set the MTU of the loopback interface
to 1500 bytes (the default value set for the ethernet
interface). \refFig{fig:ImpTSO} (top plot) shows that despite setting
the MTU to 1500 bytes, a significant number of frames have a size
greater than the MTU. Further, we observe that increasing the latency
between the peers results in a significant number of TCP segments
having a large payload length; TCP segments with large payload lengths
were also observed in frames being sent over the ethernet interface. This
increase in payload lengths is due to a feature called TCP
Segmentation Offloading (TSO)~\cite{Mogul_2003_TCPOFFLOAD}, which is
enabled by default in the 2.6 series (the current series) of the Linux
kernel. TSO enables the host machine to offload some of the TCP/IP
implementation, such as segmentation, and calculation of IP checksum,
to the network device. Further, to enhance the throughput, TSO
supports the exchange of data in frames of sizes that can be greater
that the underlying MTU size. The increase in the frame size can
result in significant improvement in throughput; however, this
improvement depends on various factors such as CPU processing power
and the amount of data being
transferred~\cite{Freimuth_2005_SERVERTSO}. \refFig{fig:ImpTSO}
(middle plot) shows the packet lengths obtained from the publicly
available traces of the Internet traffic in the WIDE
backbone~\cite{MAWI}. The values presented are from the sample taken
on the WIDE backbone on November 29, 2009. As hardware support on all
the devices in the communication link is essential for handling large
segments, the graph shows that the devices on most of links (including
end hosts or intermediate nodes) in the Internet do not support
TSO. In this paper we restrict ourselves to study the
impact of the latency, hence TSO was disabled in the subsequent
experiments. \refFig{fig:ImpTSO} (bottom plot) shows that the maximum
payload length of the frames is similar to that observed in the
Internet when the MTU is set to 1500 bytes and TSO is disabled.

The impact of the homogeneous latency and heterogeneous latency
between the peers are presented in the subsequent sections. The plots
presented are the outcome of 10 iterations.

\section{Homogeneous Latency}
\label{sec:Homogeneous}

We now present the impact of homogeneous network latency between any
two peers in a torrent on the download completion time of a file.

\begin{figure}
\begin{centering}
\includegraphics[width=0.85\columnwidth]{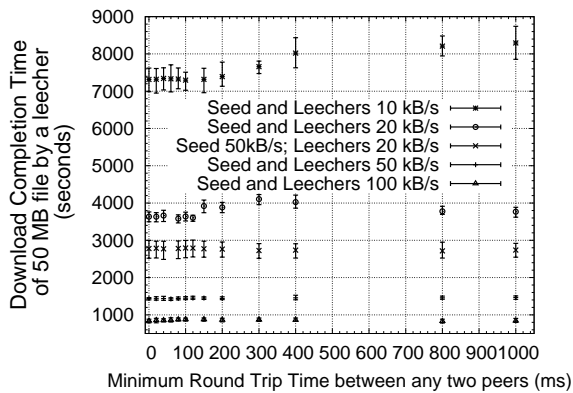}
\caption{Impact of latency on the Download Completion Time. RTT values
  mentioned are twice the latency added on a link using Netem. The
  latency increases the download completion time by at most $15\%$.}
\label{fig:SymmetricDownloadCompletion}
\vspace{-0.15in}
\end{centering}
\end{figure}

\begin{figure}
\begin{centering}
\subfloat[][Distribution of the inter-arrival time of
messages at the initial seed when RTT between peers is
0~ms.]{\label{fig:Seed0}\includegraphics[width=0.45\columnwidth]{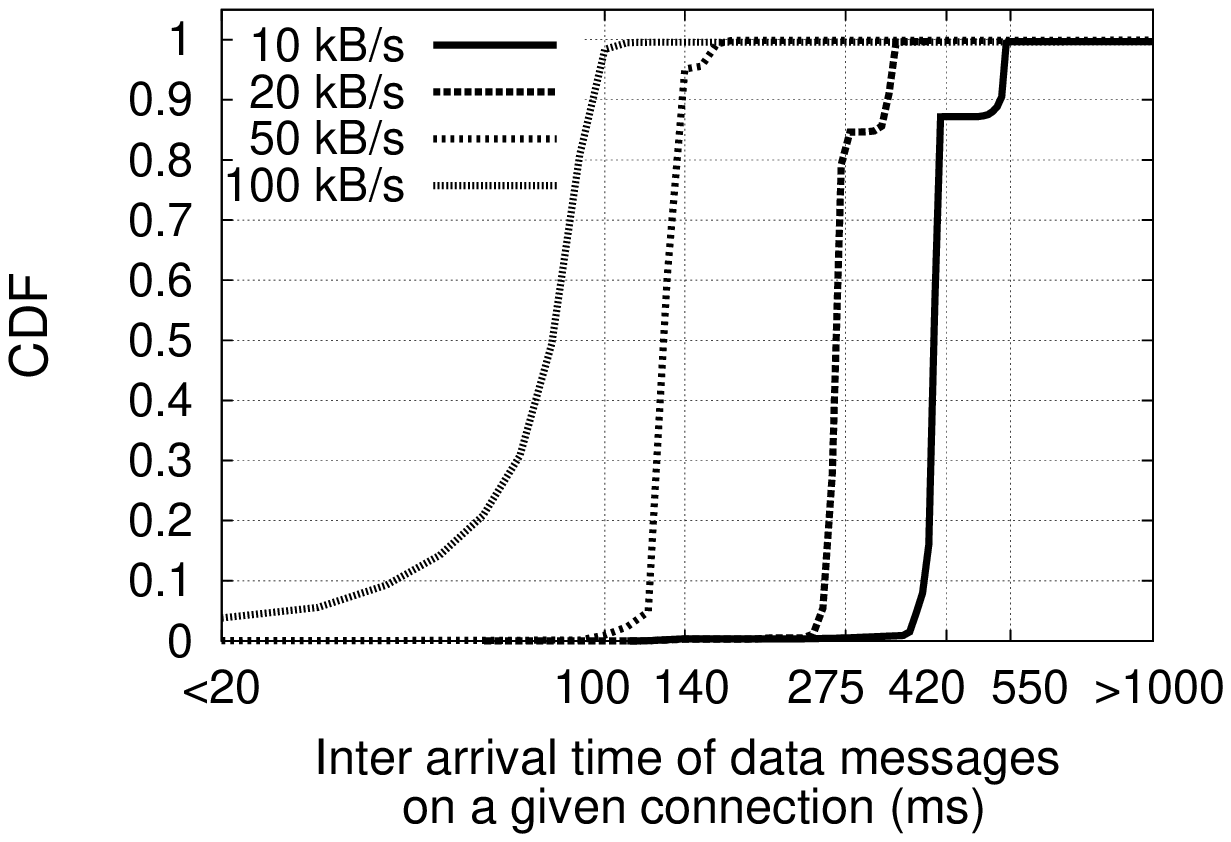}}
\hspace{0.05\columnwidth}
\subfloat[][Distribution of the inter-arrival time of
messages at the leechers when RTT between peers is 0~ms.]{\label{fig:Leecher0}\includegraphics[width=0.45\columnwidth]{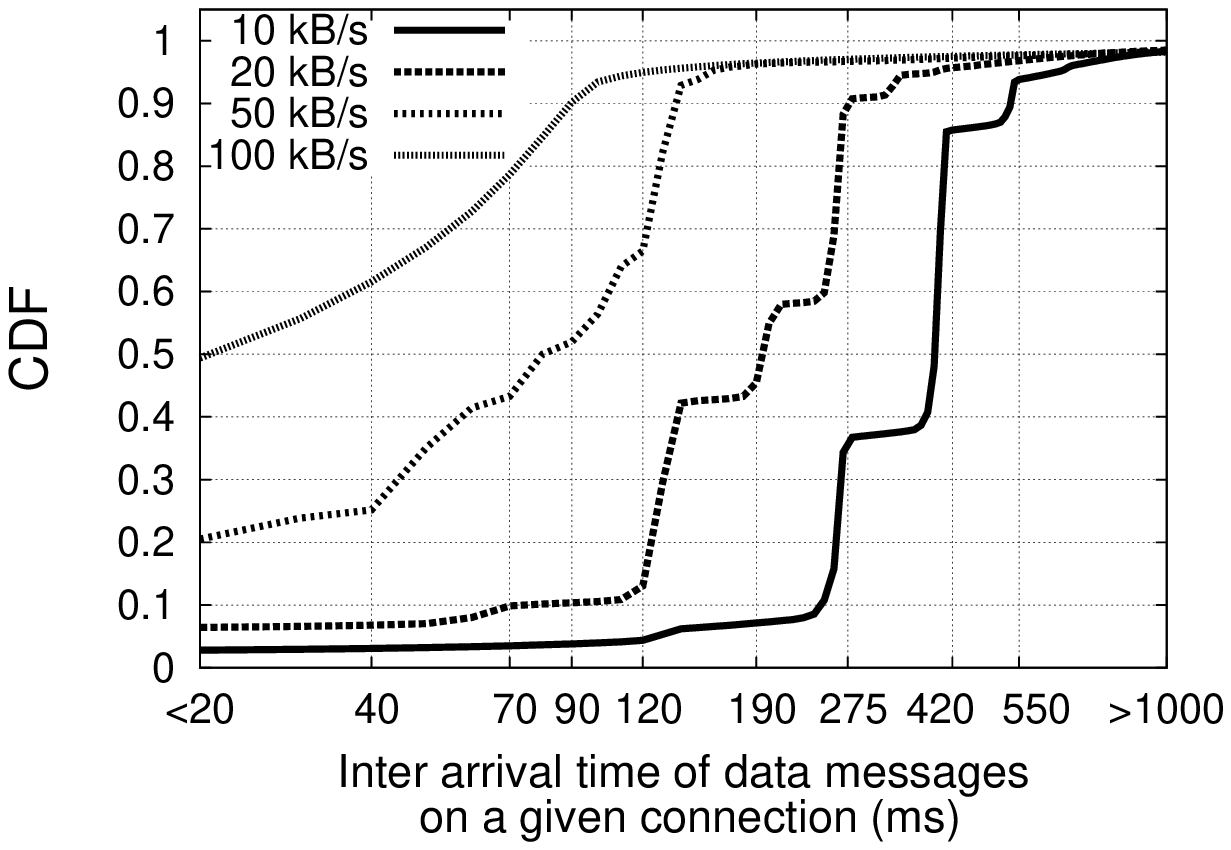}}
\caption{Distribution of the inter-arrival time of data messages when
  the RTT between the peers is 0~ms.}
\label{fig:IAT0ms}
\vspace{-0.05in}
\subfloat[][Distribution of the inter-arrival time of
messages at the initial seed when RTT between peers is
400~ms.]{\label{fig:Seed400}\includegraphics[width=0.45\columnwidth]{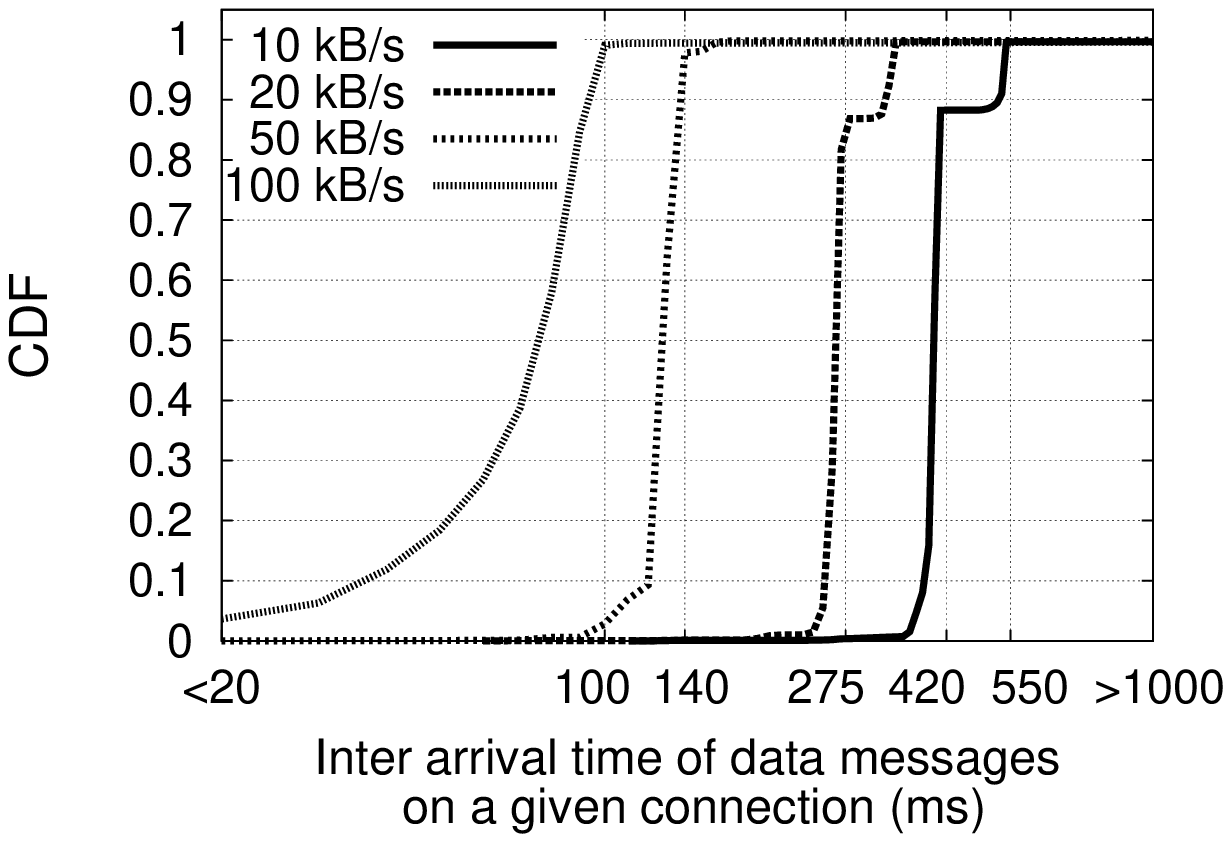}}
\hspace{0.05\columnwidth}
\subfloat[][Distribution of the inter-arrival time of
messages at the leechers when RTT between peers is 400~ms.]{\label{fig:Leecher400}\includegraphics[width=0.45\columnwidth]{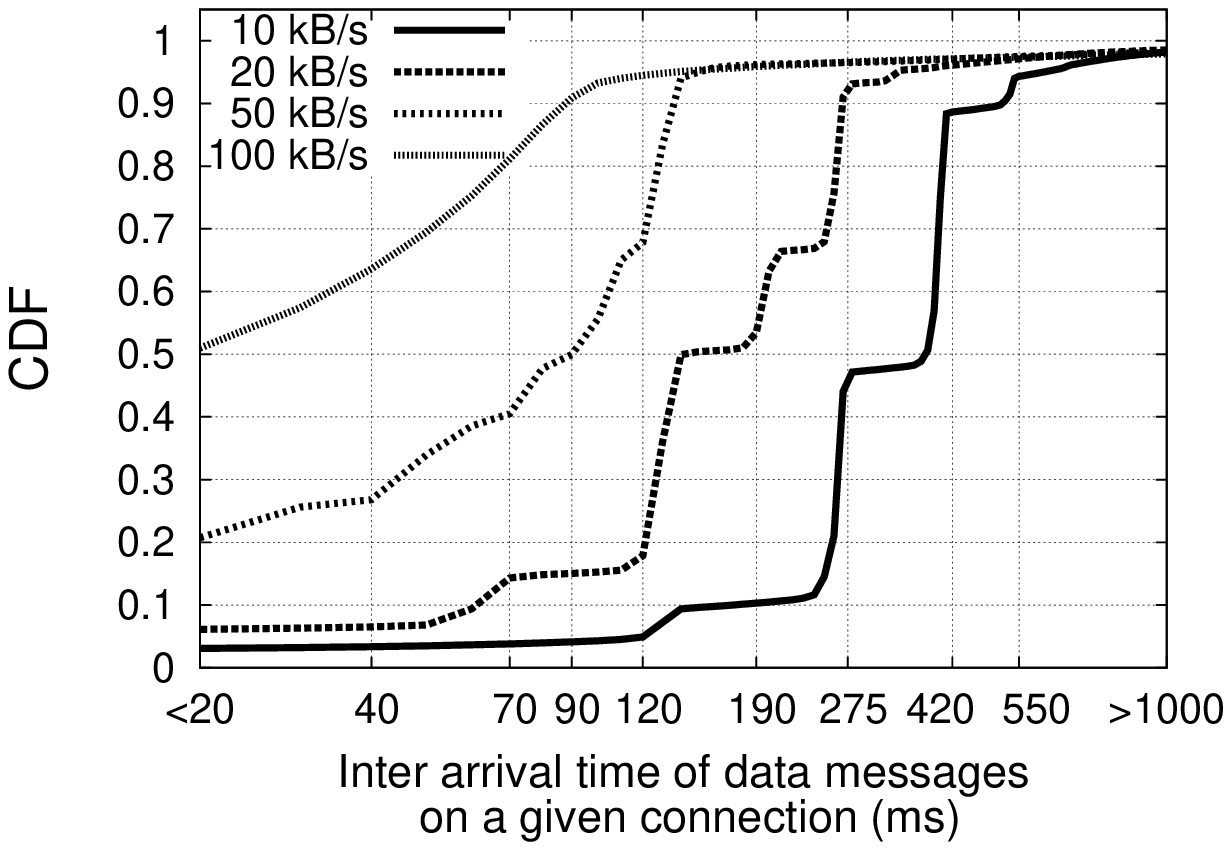}}
\caption{Distribution of the inter-arrival time of data messages when
  the RTT between the peers is 400~ms.}
\label{fig:IAT400ms}
\vspace{-0.05in}
\includegraphics[width=0.5\columnwidth]{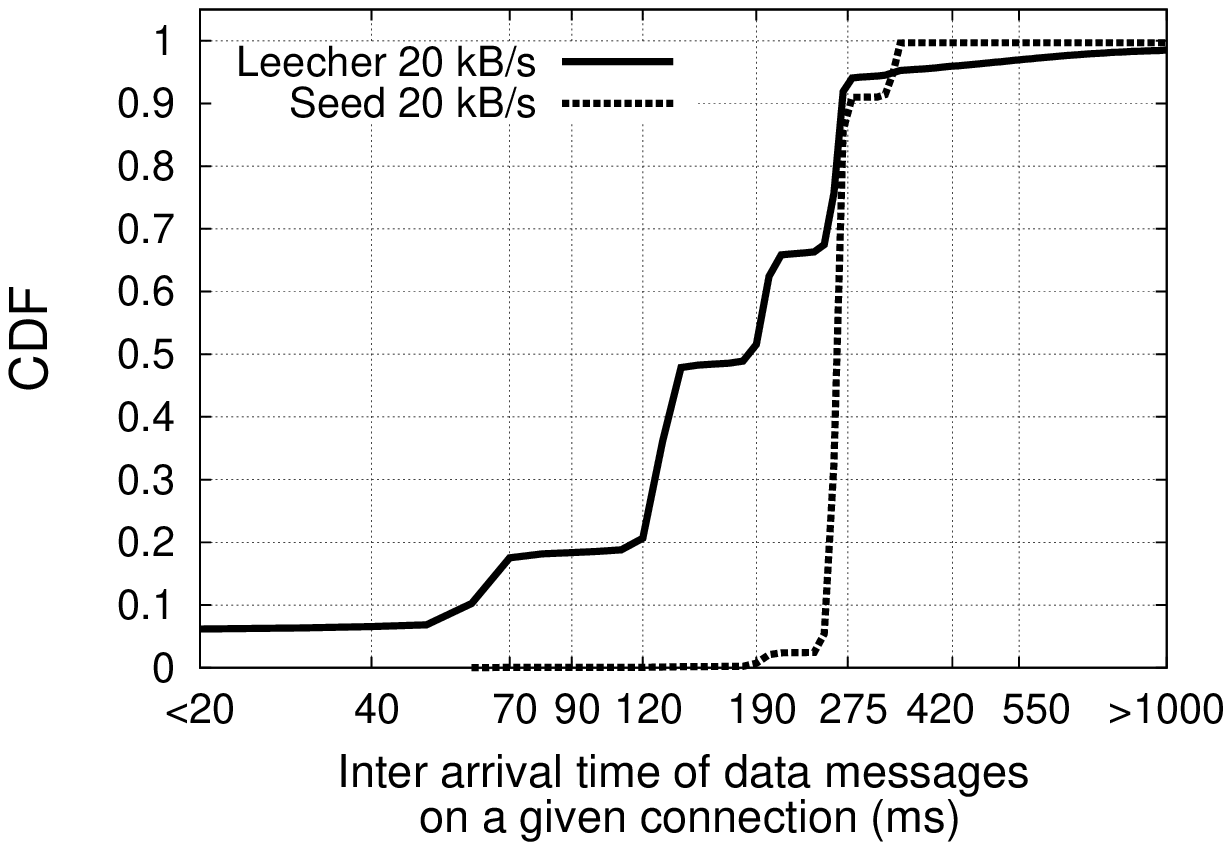}
\caption{Distribution of the inter-arrival time of
messages at the initial seed and leechers when RTT between peers is
1000~ms.}
\label{fig:IAT20kBs1000ms}
\vspace{-0.15in}
\end{centering}
\end{figure}

\subsection{Presentation of Results}

When the maximum upload rate of the seed and leechers is limited to
10~kB/s, \refFig{fig:SymmetricDownloadCompletion} shows that an RTT
greater than 200~ms results in the average download completion time
increasing by at most $15\%$. Further, when the upload rate of all the
peers is limited to 20~kB/s we observe that the download completion
time is not a monotonously increasing function of the RTT; peers
having an RTT of 1000~ms have a lower download completion time
compared to peers having an RTT of 300~ms.

However, when the maximum upload rate of the seed is increased to
50~kB/s while that of the leechers is limited to 20 kB/s, we observe
that an RTT of even 1000~ms between the peers has a
marginal impact on the download completion time of the
file. A similar observation is made when the limit on maximum upload
rates of all the peers is increased to 50~kB/s; this is also true when
the upload rates are limited to 100 kB/s.

\refFig{fig:IAT0ms},~\refFig{fig:IAT400ms},
and~\refFig{fig:IAT20kBs1000ms} show the distribution of the time
between successive \verb,send, system calls while uploading the blocks
(and pieces) of the file being
distributed.~\refFig{fig:Seed0},~\refFig{fig:Seed400}, and
\refFig{fig:IAT20kBs1000ms} show that the distribution at the seed is
similar for an RTT of 0~ms, 400~ms, and 1000~ms; further, from
~\refFig{fig:Leecher0} and~\refFig{fig:Leecher400}, we observe that
the distribution at the leechers for upload rates of 50~kB/s and
100~kB/s is the similar when the RTT is 0~ms and 400~ms. This shows
that the RTT has a marginal impact on the upload process at the peers when
their maximum upload rates are limited to 50~kB/s and
100~kB/s. However, for the upload rates of 10~kB/s and 20~kB/s we
observe that the leechers tend to have a smaller time between
successive \verb,send, system calls when the RTT is 400~ms (or,
1000~ms) compared to an RTT of 0~ms. As the peers can simultaneously
upload to many peers in parallel, the low inter-arrival time implies
that the upload capacity is being utilized to upload data to a smaller
number of peers.

\subsection{Discussion of Results}

For upload rates of 10~kB/s and 20~kB/s,
\refFig{fig:SymmetricDownloadCompletion}, \refFig{fig:IAT0ms}, and
\refFig{fig:IAT400ms}, show that when the time between successive
\verb,send, system calls is less than the RTT, the RTT does not have
an impact on the download completion time. Further, we observe that
the ramp-up period required to attain a throughput equal to the upload
rate of 50~kB/s (or 100~kB/s) does not have an impact on the download
completion time. However, we currently do not have an accurate reason
for the non-monotonous increase in the download completion time for
upload rates of 20~kB/s.

The above results show that network latency has a negligible impact
on the download completion time of a file if the peers are fast
(capable of uploading at high rates such as 50 kB/s). However, we
observe that the latency affects the download completion time when
the peers are slow (upload rates are less than or equal to
20~kB/s). Further, we observe that a single fast seed is capable of
mitigating the impact of network latency on a torrent consisting of
slow leechers.

\section{Heterogeneous Latency}
\label{sec:Heterogeneous}

The latency between two peers in a given AS, is usually less than
the latency between a peer from the given AS and another peer present
in an adjacent AS. We emulated ASes by ensuring that the latency on
the loopback interface is less than that on the ethernet interface
on each of the machines used; hence, two peers running on a given
machine have an RTT less than the RTT between a peer running on the
given machine and another peer running on another machine. Further, we
assume all the ASes to be fully meshed.

\subsection{Abstraction of ASes}

\begin{table}
\begin{footnotesize}
\begin{centering}
\begin{tabular}{|c|c|c|}
\hline
AS & Latency over & Latency over \\
   & Loopback (ms) & Ethernet (ms)  \\
\hline
$AS_{1}$ & 2 & 5 \\
\hline
$AS_{2}$ & 5 & 15 \\
\hline
$AS_3$ & 10 & 25 \\
\hline
$AS_4$ & 25 & 100 \\
\hline
$AS_5$ & 50 & 100\\
\hline
\end{tabular}
\caption{Latency values for the \emph{ingress} and \emph{egress} of
  the loopback and ethernet device while emulating an AS on a machine.}
\label{tab:LatencyAS}
\end{centering}
\end{footnotesize}
\end{table}

\begin{table}
\begin{footnotesize}
\begin{centering}
\begin{tabular}{|c|c|c|c|c|c|}
\hline
     & $AS_{1}$ & $AS_{2}$ & $AS_3$ & $AS_4$ & $AS_5$\\
\hline
$AS_{1}$ & 8 ms  & 40 ms  &   60 ms &  210 ms & 210 ms\\
\hline
$AS_{2}$ & 40 ms & 20 ms  &   80 ms &  230 ms & 230 ms\\
\hline
$AS_3$ & 60 ms  & 80 ms  &   40 ms &  250 ms & 250 ms \\
\hline
$AS_4$ & 210 ms & 230 ms &  250 ms&  100 ms & 400 ms\\
\hline
$AS_5$ & 210 ms & 230 ms &  250 ms &  400 ms & 200 ms\\
\hline
\end{tabular}
\caption{RTT between a pair of leechers. RTT between a leecher in $AS_1$
  and a leecher in $AS_5$ is 210 ms.}
\label{tab:RTTAS}
\end{centering}
\end{footnotesize}
\end{table}

\begin{table}
\begin{footnotesize}
\begin{centering}
\begin{tabular}{|c|c|c|c|c|c|}
\hline
     & $AS_{1}$ & $AS_{2}$ & $AS_3$ & $AS_4$ & $AS_5$\\
\hline
$AS_{1}'$ & 20 ms  & 40 ms  &   60 ms &  210 ms & 210 ms\\
\hline
$AS_{2}'$ & 40 ms & 60 ms  &   80 ms &  230 ms & 230 ms\\
\hline
$AS_3'$ & 60 ms  & 80 ms  &   100 ms &  250 ms & 250 ms \\
\hline
$AS_4'$ & 210 ms & 230 ms &  250 ms&  400 ms & 400 ms\\
\hline
$AS_5'$ & 210 ms & 230 ms &  250 ms &  400 ms & 400 ms\\
\hline
\end{tabular}
\caption{RTT between the initial seed and the other peers
  (except the tracker) in the torrent. RTT between the seed in $AS_1'$
  and a peer in $AS_1$ is 20 ms.}
\label{tab:RTTASSeed}
\end{centering}
\end{footnotesize}
\end{table}

As in the case of homogeneous latency, we consider a torrent
consisting of three hundred leechers, one tracker, and one initial
seed; we emulated three ASes with one hundred leechers each, while the
seed and the tracker were placed in the fourth AS. The four ASes used
in these experiments were chosen from a set five ASes; the latency
values added on the \emph{ingress} and \emph{egress} of the loopback
and ethernet device of the machines while emulating these five ASes are
given in \refTab{tab:LatencyAS}. We now show how
\refFig{fig:Grid5KTopology} and \refTab{tab:LatencyAS} can be used to
find the RTT between a pair of peers.

\begin{figure}
\begin{centering}
\subfloat[][Download completion time for leechers present in $AS_{1}$, $AS_{2}$, and
$AS_{3}$. ]{\label{fig:Asym20AS123}\includegraphics[width=0.8\columnwidth]{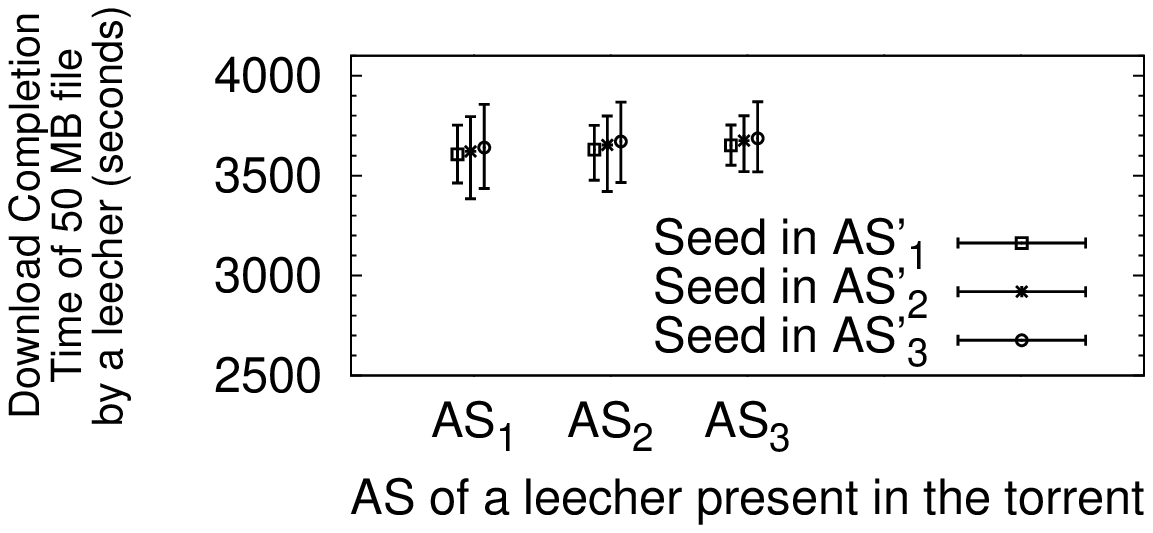}}

\subfloat[][Download completion time for leechers present in $AS_{1}$, $AS_{3}$, and
$AS_{4}$. Having peers in an AS with large latency ($AS_{4}$) and the initial seed in an AS
with large latency ($AS_{4}'$) affects the download completion time.]{\label{fig:Asym20AS134}\includegraphics[width=0.8\columnwidth]{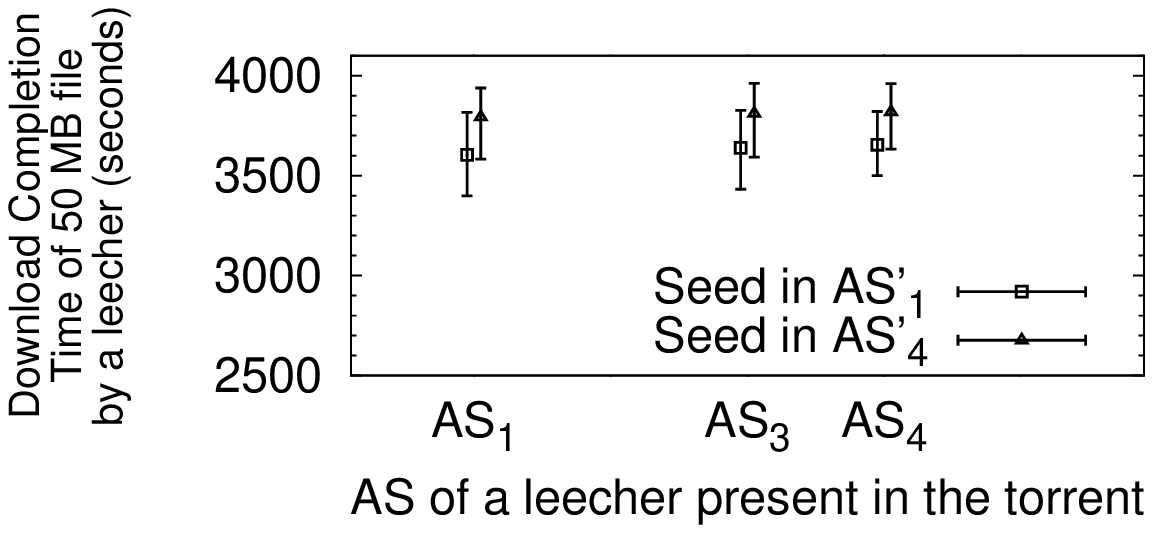}}

\subfloat[][Download completion time for leechers present in $AS_{1}$, $AS_{3}$, and
$AS_{5}$. Having peers in an AS with large latency ($AS_{5}$) and the initial seed in an AS
with large latency ($AS_{5}'$) affects the download completion time.]{\label{fig:Asym20AS135}\includegraphics[width=0.8\columnwidth]{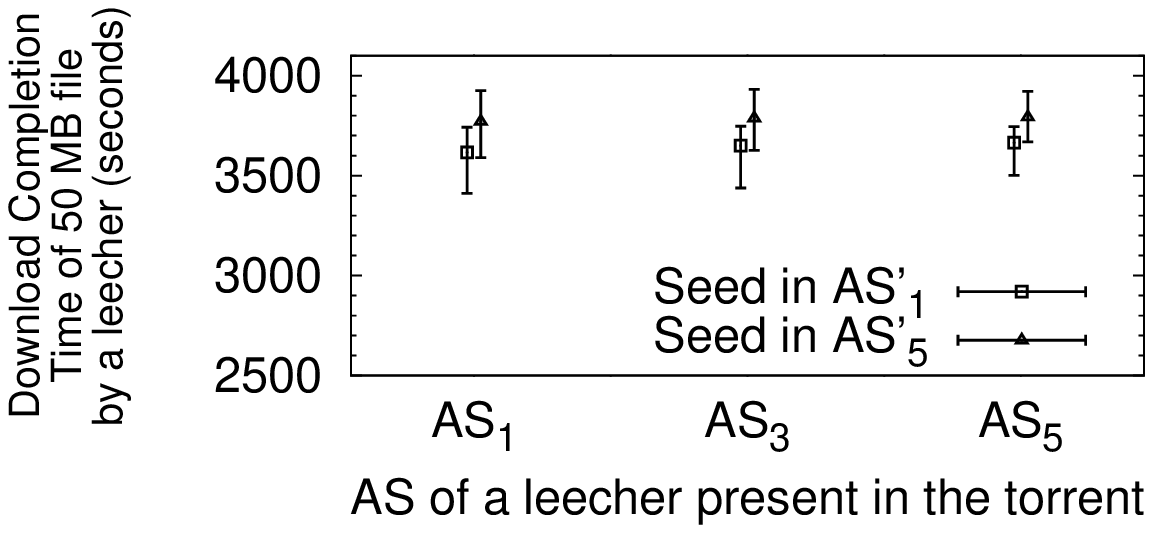}}
\caption{Download completion time of a 50 MB file by leechers in a
  given AS when the maximum upload rate of all the peers is
  20~kB/s.}
\label{fig:AsymUp20}
\includegraphics[width=0.8\columnwidth]{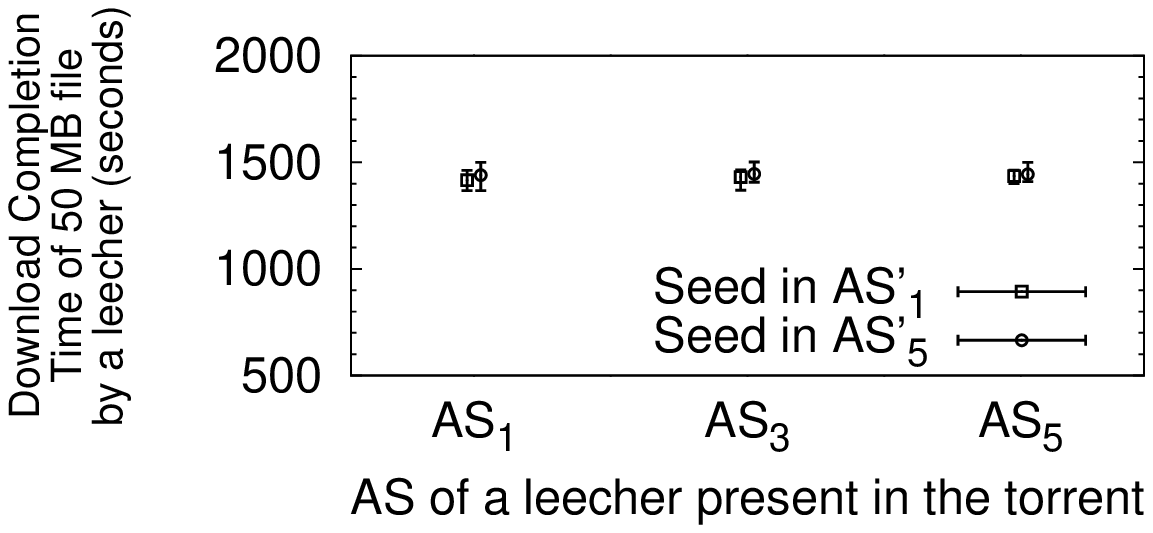}
\caption{Download completion time of a 50 MB file by leechers present
  in a given AS when the maximum upload rate of all the peers is
  50~kB/s.}
\label{fig:AsymUp50}
\end{centering}
\end{figure}

As a machine is used to emulate an AS, a peer in $AS_1$ uses the
ethernet interface to communicate with a peer in $AS_{2}$, the RTT
between this pair of peer is therefore 40~ms (5+15+15+5); as a
peer in $AS_1$ uses the loopback interface to communicate with another
peer in $AS_1$, the RTT between this pair of peers is 8~ms
(2+2+2+2). \refTab{tab:RTTAS} gives the RTT values between all such
pairs of peers that are initially leechers.

Similarly, from \refFig{fig:Grid5KTopology}, as the
initial seed (henceforth called as the seed) and leechers are placed
in different machines, the seed uses the ethernet interface of the
machine to communicate with all the leechers present in the
torrent. We use $AS_i'$ to denote that the seed
and the tracker are placed in an AS with an inter-AS latency and
intra-AS latency equal to that of $AS_i$; for example, $AS_{1}'$
implies that the seed and tracker are placed in an AS having the same
latency values as $AS_{1}$. Therefore, the RTT between the seed in
$AS_{1}'$ and a leecher in $AS_{1}$ is 20~ms ($5+5+5+5$), while
the RTT between the same seed and a leecher in $AS_{5}$ is 210~ms. The RTT
between the seed and the peers are given in \refTab{tab:RTTASSeed}.

From \refTab{tab:RTTAS}, the RTT between a peer in either $AS_{1}$,
$AS_{2}$, or $AS_3$, and another peer in either $AS_{1}$, $AS_{2}$, or
$AS_{3}$, is less than 100~ms. Further, the RTT between a peer in
$AS_{4}$ and another peer in $AS_{4}$ is 100~ms, while the RTT from
this peer to any other peer is greater than 200~ms. Similarly the RTT
from a peer in $AS_{5}$ to any other peer, irrespective of its AS, is
greater than 200~ms.

\subsection{Presentation and Discussion of Results}

\refFig{fig:AsymUp20} and \refFig{fig:AsymUp50} show the impact of
heterogeneous latency on the download completion time of a 50 MB
file when the upload rate of the peers is limited to 20~kB/s and
50~kB/s respectively. The X-axis represents the AS of the leechers
present in the torrent, and the Y-axis represents the download
completion time in seconds; the error bars indicate the minimum and
maximum download completion time of the leechers in 10 iterations.

\refFig{fig:Asym20AS123} shows the outcome of three experiments
having the leechers placed in $AS_{1}$, $AS_{2}$, and $AS_{3}$;
for the first experiment the seed was placed in $AS_{1}'$, for the other
two experiments, the seed was placed in $AS_{2}'$ and $AS_{3}'$
respectively. The figure shows that the RTT between the peers does not
affect the download completion time. According
to~\refTab{tab:RTTAS}, the RTT between any two peers in these
experiments was less than 120~ms, hence, as in the case of homogeneous
latency, we observe that an RTT of less than 120~ms does not affect
the download completion time when the upload rates are limited to
20~kB/s.

When some of the peers are present in an AS having a large RTT
($AS_{4}$ or $AS_{5}$), and the seed is also present in another AS
with a large RTT ($AS_{4}'$ or $AS_{5}'$), then
\refFig{fig:Asym20AS134}, and \refFig{fig:Asym20AS135}, show that the
RTT affects the download completion time. However, we observe that
the increase in average download completion time is not more than
$15\%$ of the average download completion time when all the peers in a
torrent have an RTT less than 120~ms.

\refFig{fig:AsymUp50} shows the impact of heterogeneous latency
on the download completion time when the maximum upload rate of all the
peers in the torrent is limited to 50~kB/s. We observe that an RTT of
400~ms, between the seed in $AS_{5}'$ and the leechers in $AS_{5}$,
does not have a significant impact on the download completion time of
the file. These observations are in line with the observations in
\refSec{sec:Homogeneous}.

\refFig{fig:SymmetricDownloadCompletion},
\refFig{fig:AsymUp20}, and
\refFig{fig:AsymUp50}, confirm that the topology of
homogeneous latency provides an upper bound on the download
completion time of a file when the maximum latency between any two
peers in a torrent is known. Further, the observations made in
\refSec{sec:Homogeneous} can be used in experiments where the
latency between a pair of peers is heterogeneous.

\section{Conclusion}
\label{sec:Conclusion}

The network latency between the peers has a marginal impact on the
download completion time when the peers have their
upload rates limited to high values such as 50~kB/s and
100~kB/s; our experiments show that the ramp-up period which is
required to attain the throughput equal to these upload rates has a
marginal impact on the download completion time. When the
peers are slow (upload rates limited to  values less than or equal to
20~kB/s) we observe that the download completion time is affected by
the network latency; however, the increase in the average download
completion time is not more than $15\%$ of the average download
completion time when there is no network latency between the
peers. As the network latency has a marginal impact on the time
required to download a file, \emph{BitTorrent experiments can be
 performed on testbeds without explicitly emulating latency between
 the peers in a torrent}.

\section{Acknowledgment}

Experiments presented in this paper were carried out using the Grid'5000
experimental testbed, being developed under the INRIA ALADDIN development
action with support from CNRS, RENATER and several Universities as well
as other funding bodies (see https://www.grid5000.fr).

{\bibliographystyle{IEEEtran}
\bibliography{IEEEabrv,biblio}}

\begin{thebibliography}{1}
\providecommand{\url}[1]{#1}
\csname url@samestyle\endcsname
\providecommand{\newblock}{\relax}
\providecommand{\bibinfo}[2]{#2}
\providecommand{\BIBentrySTDinterwordspacing}{\spaceskip=0pt\relax}
\providecommand{\BIBentryALTinterwordstretchfactor}{4}
\providecommand{\BIBentryALTinterwordspacing}{\spaceskip=\fontdimen2\font plus
\BIBentryALTinterwordstretchfactor\fontdimen3\font minus
  \fontdimen4\font\relax}
\providecommand{\BIBforeignlanguage}[2]{{%
\expandafter\ifx\csname l@#1\endcsname\relax
\typeout{** WARNING: IEEEtran.bst: No hyphenation pattern has been}%
\typeout{** loaded for the language `#1'. Using the pattern for}%
\typeout{** the default language instead.}%
\else
\language=\csname l@#1\endcsname
\fi
#2}}
\providecommand{\BIBdecl}{\relax}
\BIBdecl

\bibitem{Cohen_2008_BitTorrentProtocol}
B.~Cohen, \emph{{The BitTorrent Protocol Specification}}, Jan 2008.

\bibitem{Mathis_1997_MBTCP}
M.~Mathis, J.~Semke, J.~Mahdavi, and T.~Ott, ``{The Macroscopic Behavior of the
  TCP Congestion Avoidance Algorithm},'' \emph{SIGCOMM Computer Communication
  Review}, vol.~27, no.~3, pp. 67--82, 1997.

\bibitem{grid5000}
``https://www.grid5000.fr.''

\bibitem{Hemminger_2005_NETEM}
S.~Hemminger, ``{Network emulation with NetEm},'' in \emph{Linux Conf Au},
  2005.

\bibitem{Mogul_2003_TCPOFFLOAD}
J.~C. Mogul, ``Tcp offload is a dumb idea whose time has come,'' in
  \emph{HotOS}, 2003, pp. 25--30.

\bibitem{Freimuth_2005_SERVERTSO}
D.~Freimuth, E.~Hu, J.~LaVoie, R.~Mraz, E.~Nahum, P.~Pradhan, and J.~Tracey,
  ``{Server Network Scalability and TCP offload},'' in \emph{ATEC '05:
  Proceedings of the USENIX Annual Technical Conference}.\hskip 1em plus 0.5em
  minus 0.4em\relax Berkeley, CA, USA: USENIX Association, 2005, pp. 15--15.

\bibitem{MAWI}
``http://mawi.nezu.wide.ad.jp/mawi/.''

\end{thebibliography}

\end{document}